# EDEPR of impurity centers embedded in silicon microcavities

N.T. Bagraev,[a,*] W. Gehlhoff,[b] D.S. Gets,[a] L.E. Klyachkin,[a] A.A. Kudryavtsev,[a]

A.M. Malyarenko,[a] V.A. Mashkov,[a] V.V. Romanov,[a]

[a]*Ioffe Physical-Technical Institute, 194021, St.Petersburg, Russia*

[b]*Institut für Festkörperphysik, TU Berlin, D-10623 Berlin, Germany*

**Abstract**

We present the first findings of the new electrically-detected EPR (EDEPR) technique which reveal different shallow and deep centers without using the external cavity as well as the hf source and recorder, with measuring the only magnetoresistance of the Si-QW confined by the superconductor δ-barriers.

[*] Corresponding author. Tel.: +7-812-292-73-15; fax: +7-812-297-10-17; e-mail: impurity.dipole@mail.ioffe.ru

## 1. Introduction

Spin-dependent transport through semiconductor nanostructures embedded in nano- and microcavities between superconductor leads is of great interest to identify the magnetic resonance phenomena without using both the external cavity and the external hf sources and recorders [1, 2]. One of the best candidate on the role of such a 'sandwich' structure that is able to demonstrate the electrically- and optically-detected EPR (EDEPR, ODMR) by measuring the only magnetoresistance at high temperature appears to be the high mobility p-type silicon quantum well (Si-QW), 2nm, confined by the δ-barriers heavily doped with boron (Fig. 1). The findings of the electrical resistivity, thermo-emf and magnetic susceptibility measurements are actually evidence of the superconductor properties for these δ-barriers, 3 nm, N(B)=5 $10^{21}$ cm$^{-3}$, which are revealed at critical sheet density of holes, > $10^{11}$ cm$^{-2}$, in the p-type Si-QW on the n-type Si (100) surface [3]. These silicon nanostructures embedded in superconductor shells have been shown to be type II high temperature superconductors (HTS) with $T_c$=145 K and $H_{c2}$=0.22 T [2]. Here the S-Si-QW-S structures performed in the Hall geometry are used to register the EPR spectra of single impurity centers using the EDEPR technique above noticed.

## 2. Methods

The energy positions of the two-dimensional subbands of holes in the Si-QW and the value of the superconductor gap, 2Δ=0.044 eV, caused by the superconductor δ-barriers were determined in the studies of the far-infrared and tunneling spectroscopy [2, 4]. The superconductor gap appeared to be the source of the THz emission due to the Josephson transitions self-assembled in the sandwich structure (Fig.1). Besides, the EPR studies have revealed that the δ-barriers appear to consist of the trigonal dipole centers, $B^+$ - $B^-$, which are caused by the negative-U reconstruction of the shallow boron acceptors, $2B^0$=>$B^+$ + $B^-$ (Figs. 2a and 2b) [3, 4]. The high temperature superconductor properties for the δ-barriers have been shown to result from the transfer of the small hole bipolarons through these negative-U dipole centers of boron, which cause the GHz generation under applied voltage, optical pumping or by scanning external magnetic field with the enhancement by varying the positions of the leads within frameworks of the Hall geometry (Fig. 1) [3, 4]. Spectroscopic studies of the Rabi splitting in the GHz range carried out with the

X-band have confirmed this pattern and furthermore demonstrated that the area of the δ-barrier defines the dimensions of the GHz cavity incorporated into the Si-QW. Thus, the sandwich structure provides the GHz generation thereby giving rise to the EDEPR measurements of single impurity centers without using the hf source and recorder.

## 3. Results

The phosphorus EPR lines with the characteristic hf splitting of 4.1 mT are observed, with a complicated behavior of intensities and phases due to effects of a spin-dependent scattering (Fig. 3). Besides, the spin-dependent scattering of 2D holes on the phosphorus shallow centers is revealed by measuring the phosphorus line splitting that is evidence of the exchange interaction, which is similar to the effect of zero-field splitting in one-dimensional channels [3]. It should be noted that this considerable splitting of the P-lines has been found, for the first time, in the same device using the ordinary EDEPR technique [5]. The high sensitivity of the new EDEPR technique is confirmed by the measurements of the NL8 spectrum that identifies residual oxygen thermodonors, $TD^+$ state, in the p-type Si-QW (Fig. 3) [6]. This center of the orthorhombic symmetry has been also found by the ordinary EDEPR method in the sandwich structure discussed here [7]. The central lines in the EDEPR spectrum shown in third figure are different a little from the NL10 spectrum that is related to the neutral thermodonor containing a single hydrogen atom. Nevertheless, this EDEPR spectrum appears to identify the hydrogen-related center in the p-type Si-QW, because its characteristic hf splitting, 23 MHz, corresponds to the hf hydrogen splitting [8]. Different phase of the hf lines is of importance to be noticed, which seems to result from the high spin polarization.

The 23 MHz hf splitting is verified also in the EDEPR line with a g-value of 2.07 like $Fe^0$ in bulk silicon (Fig. 4a). This EDEPR spectrum is of interest to exhibit a strong angular dependence of the line intensity with maximum for B||<111> that is practically the same as for the iron-related center identified by the ordinary EDEPR, with the observation of the double quantum transitions (Fig. 4b) [5]. The EDEPR spectrum shown in figure 4b is not related however to the well-known FeH center [9] and seems to be as a result of the hydrogen passivation of interstitial $Fe^0$ center.

The high sensitivity of the EDEPR technique demonstrated allows the studies in weak magnetic fields that are of importance for the measurements of the hf splitting for the centers inserted in the quantum wells, which are characterized by the large g-values. Firstly, this advantage is revealed by measuring the EDEPR spectrum of the $Fe^+$ center, which appears to exhibit the hf $^{29}Si$ splitting in the absence of the external cavity as well as the hf source and recorder (Fig. 5).

Secondly, the hf structure of the erbium-related center is found, for the first time, in silicon (Fig. 6). Erbium doping was done at the diffusion temperature of $1100^0C$ in the process of long-time diffusion accompanied by surface injection of vacancies from the interface between the oxide overlayer and the n-type Si (100) substrate. Then, the sandwich structure that represents the p-type Si-QW confined by the superconductor δ-barriers was prepared on the n-type Si (100) (Fig. 1). Thus, small concentration of the erbium-related centers is a basis of the EDEPR and ODMR observation that are caused by the spin-dependent scattering of 2D holes. The g-value, 4.82, and the trigonal symmetry of the erbium-related center identified from the angular dependences of the EDEPR spectrum observed are evidence of its similarity to the erbium center studied by the ordinary EPR [10]. The components of the hf erbium structure, $I=7/2$, are seen to be split in four lines (Fig. 6). This splitting seems to result from the hf structure of boron ($I=3/2$) that forms the trigonal dipole centers in the δ-barriers. The results obtained allow the model of the erbium-related center in the p-type Si-QW within frameworks of the replacement of one boron atom in the trigonal dipole center by erbium thereby forming single dipole centers $B^+$-$Er^-$ (Fig. 7a). The paramagnetic state of this center seems to be created by the capture of 2D holes (Fig. 7b) transferred along edged channels in the sandwich structures.

## 4. Summary

The new electrically-detected EPR (EDEPR) technique, which allows the studies without using the external cavity and the hf source and recorder, has been demonstrated by measuring the only magnetoresistance. Using this new EDEPR technique, the EPR spectrometer has been suggested to be replaced by the sandwich structure that represents the ultra-narrow silicon quantum well confined by the superconductor δ-barriers. These δ-barriers appeared to be the sources of the GHz emission that can be enhanced by varying the dimensions of the sandwich structure which is able to form the internal cavity. This new EDEPR technique has been applied to the studies of the phosphorus, iron-, hydrogen- and erbium-related centers embedded in the ultra-narrow p-type silicon quantum well that was a basis of the sandwich structure. The hf erbium structure has been found, for the first time, for the erbium-related centers in silicon.

**5. Acknowledgements**

The work was supported by the SNSF programme (grant IB7320-110970/1), RAS-QM (grant P-03. 4.1).

**References**

1. N.T. Bagraev et al., Physica B **340-342** (2003) 1078.

2. N.T. Bagraev et al., Physica C **468** (2008) 840.

3. N.T. Bagraev et al., J. Phys.:Condens. Matter, **20** (2008) 164202.

4. N.T. Bagraev et al., Physica C **437-438** (2006) 21.

5. W. Gehlhoff et al., Materials Sci.Forum **196-201** (1995) 467.

6. H.H.P.Th. Bekman et al., Phys. Rev. Lett. **61** (1988) 227.

7. W. Gehlhoff et al. in Shallow Level Centres in Semiconductors, edited by C.A.J. Ammerlaan and B. Pajot, World Scientific, Singapore, 1997, p. 227

8. Yu.V. Gorelkinskii et al., Physica B **170** (1991) 15.

9. C.A.J. Ammerlaan et al., Solid State Phenom. **85-86** (2002) 353.

10. J.D. Carey et al. Phys. Rev. B **59** (1999) 2773

**Captions**

Fig. 1. The schematic diagram of the experimental device that demonstrates a perspective view of the p-type Si-QW confined by the δ-barriers heavily doped with boron on the n-type Si (100) surface. The top gate is able to control the sheet density of holes and the Rashba SOI value. The depletion regions indicate the Hall geometry of leads.

Fig. 2. (a) Model for the elastic reconstruction of a shallow boron acceptor which is accompanied by the formation of the trigonal dipole ($B^+$ - $B^-$) centers as a result of the negative-U reaction: $2B^o \rightarrow B^+ + B^-$. (b) A series of the dipole negative-U centers of boron located between the undoped microdefects that seem to be a basis of nanostructured δ - barriers confining the Si-QW.

Fig 3. EDEPR of phosphorus, NL8 and hydrogen-related centers in the Si-QW confined by the superconductor δ-barriers, which is observed by measuring a magnetoresistance without the external cavity as well as the hf source and recorder. T=77 K. B∥<100> in the plane ⊥ to the {100} interface. ν=9.08 GHz. The 23 MHz splitting revealed by the central lines seems to be evidence of the hf hydrogen structure.

Fig. 4. (a) EDEPR of a FeH-related center in the Si-QW confined by the superconductor δ-barriers, which is observed by measuring a magnetoresistance without the external cavity as well as the hf source and recorder. T=77 K. B∥<100> in the plane ⊥ to the {100} interface. ν=9.08 GHz. The 23 MHz splitting seems to be caused by the HFI with hydrogen.

(b) The hf structure of the X-line observed with ordinary EDEPR at 50mW and low modulation amplitude (0.05mT); T=3.7K, B ∥ <111>, 50 scans [5].

Fig. 5. EDEPR of the $Fe^+$ center in the Si-QW confined by the superconductor δ-barriers, which is observed by measuring a magnetoresistance without the external cavity as well as the hf source and recorder. T=77 K. B∥<100> in the plane ⊥ to the {100} interface. ν=9.08 GHz.

Fig. 6. EDEPR of the trigonal Er-related center in the Si-QW confined by the superconductor δ-barriers, which is observed by measuring a magnetoresistance without the external cavity as well as the hf source and recorder. T=77 K. B∥<100> in the plane ⊥ to the {100} interface. ν=9.08 GHz. The hf erbium

structure (I=7/2) are split in four lines that seem to be due to the presence of boron (I=3/2) inside of the Er-related center.

Fig. 7. Model for the trigonal dipole boron-erbium center (a), with paramagnetic state created by the capture of 2D holes in the Si-QW confined by the superconductor δ-barriers (b).

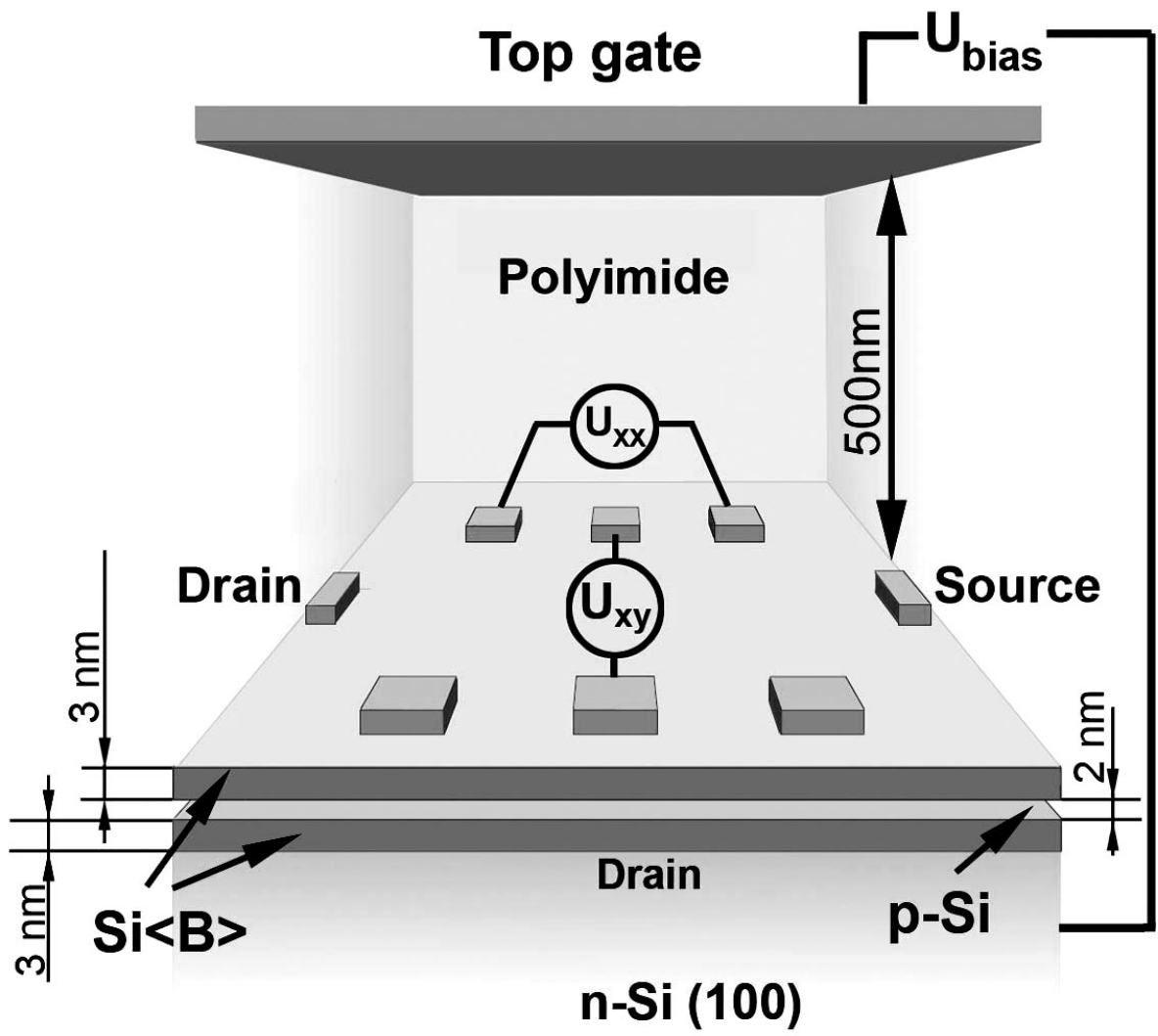

Fig.1.

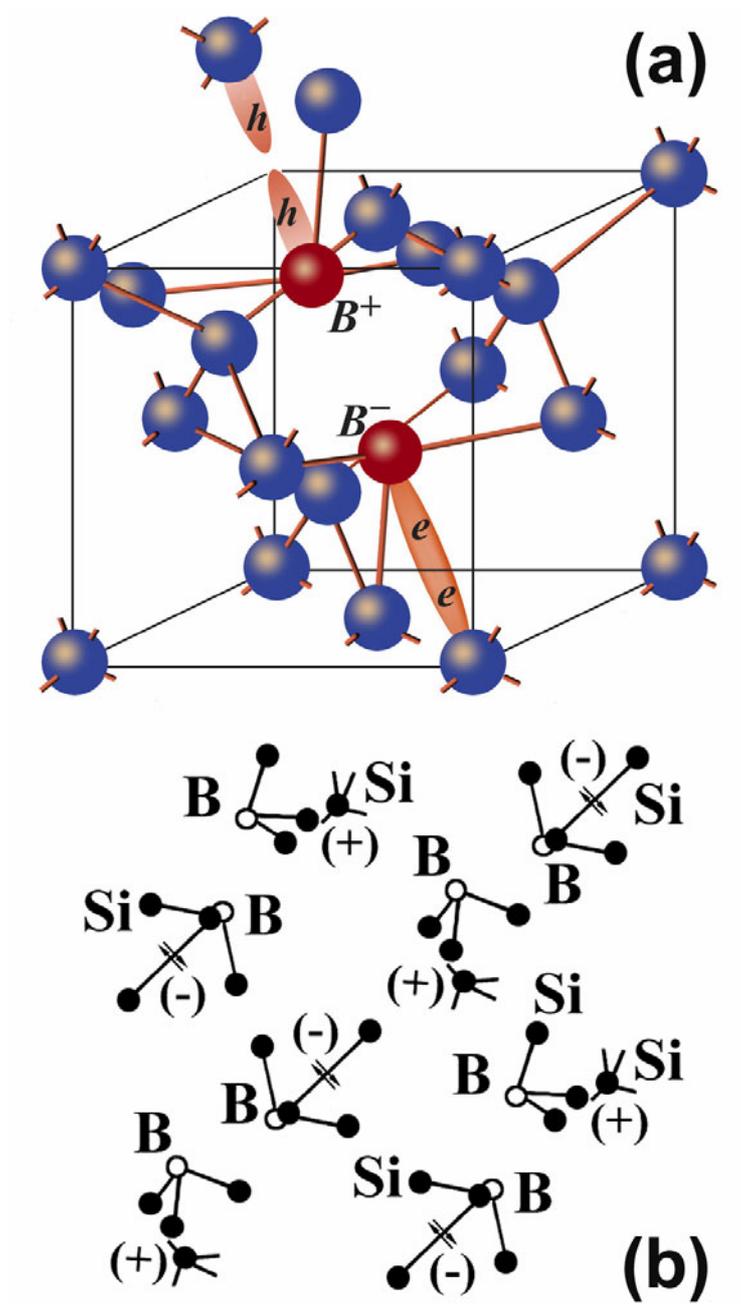

Fig. 2.

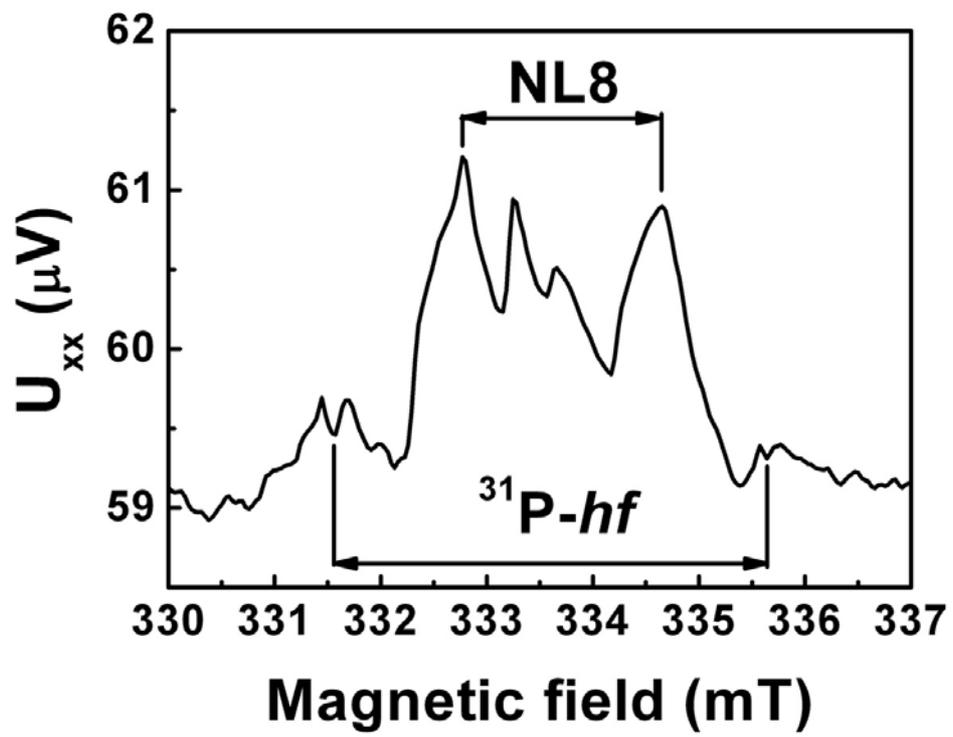

Fig. 3.

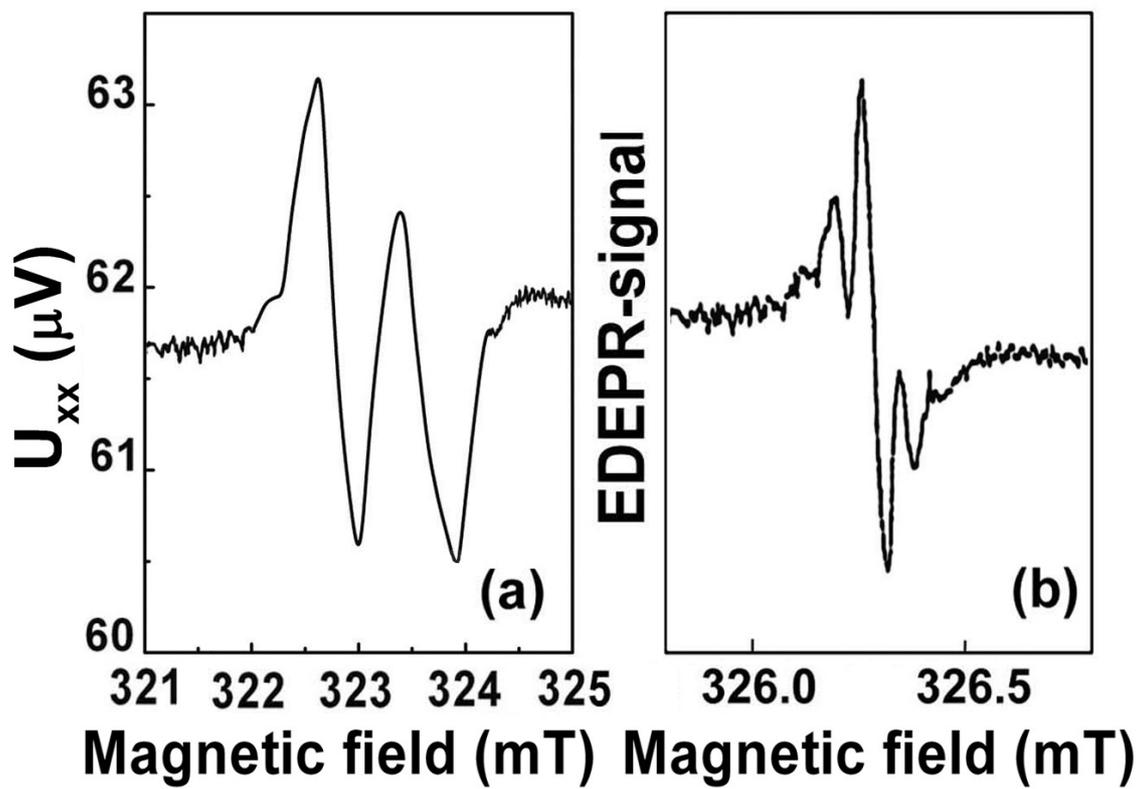

Fig. 4.

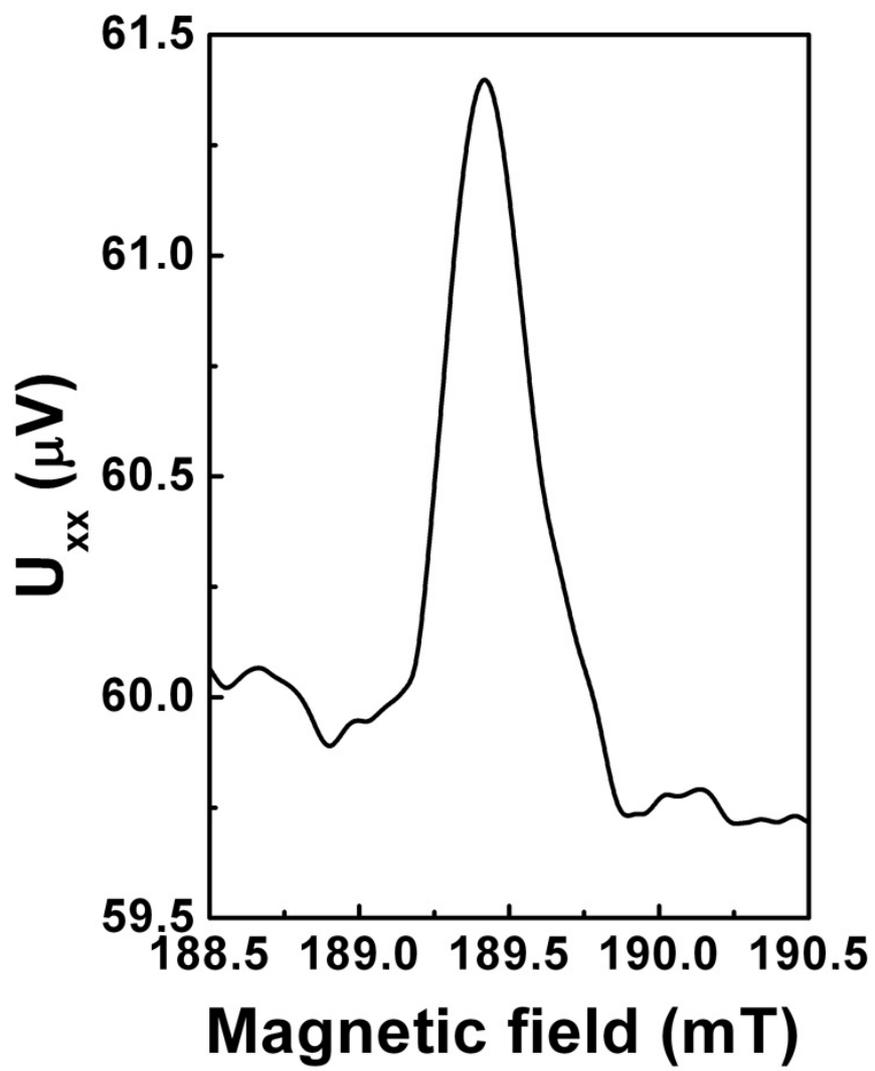

Fig. 5.

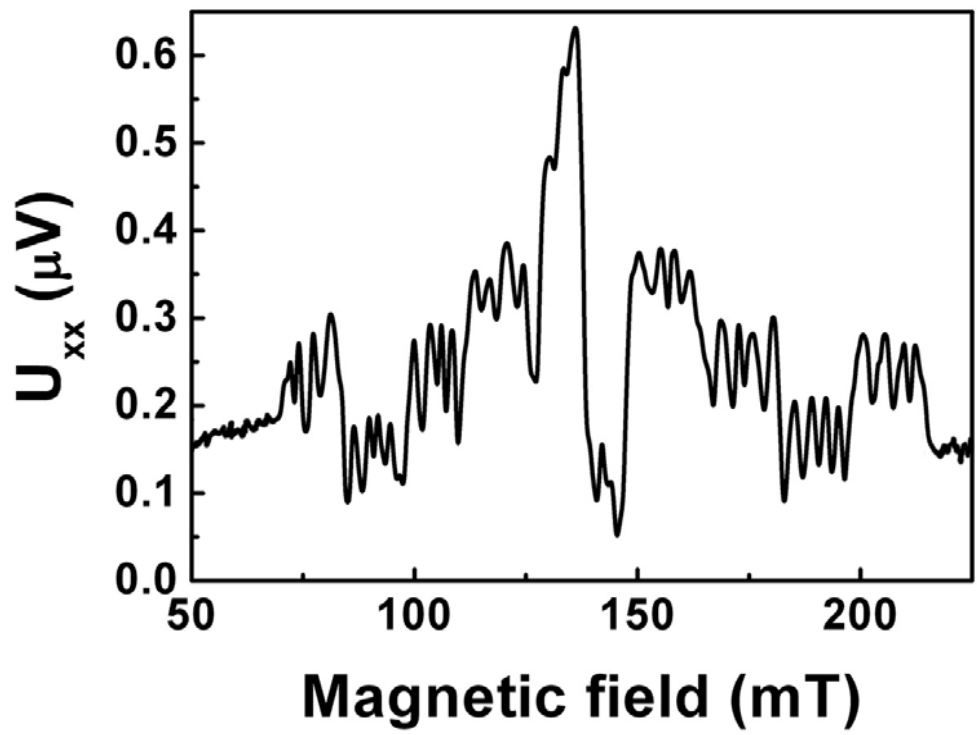

Fig. 6.

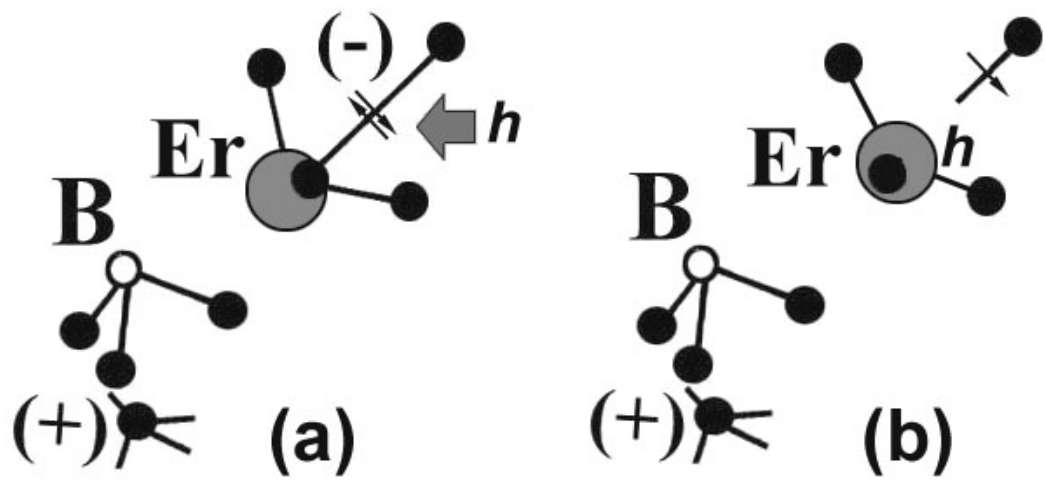

Fig. 7.